\documentclass[10pt,letterpaper,twocolumn]{article}
\usepackage{ol2}
\usepackage{eurosym}
\usepackage[draft]{hyperref}
\usepackage{amsmath}
\usepackage{amssymb}
\usepackage{color}

\newcommand{\PT}{{\cal PT}}

\begin{document}

\twocolumn[

\title{Families of stationary modes in complex potentials
}
\author{Vladimir V. Konotop %$^{1}$
and  Dmitry A. Zezyulin%$^{1}$
}

\address{
Centro de F\'{\i}sica Te\'orica e Computacional,
Faculdade de Ci\^encias, Universidade de Lisboa,  Avenida
Professor Gama Pinto 2, Lisboa 1649-003, Portugal
\\ Departamento de F\'isica, Faculdade de Ci\^encias, Universidade de Lisboa,  Campo Grande, Ed. C8,
Lisboa 1749-016, Portugal
}

\begin{abstract}
It is shown that a general class of complex asymmetric potentials of the form $ w^2(x)-iw_x(x)$, where $w(x)$ is a real function, allows for the existence of one-parametric continuous families of the stationary nonlinear modes bifurcating from the linear spectrum and propagating in Kerr media. As an example, we introduce an asymmetric double-hump complex potential and show that it supports continuous families of stable  nonlinear modes.
\end{abstract}


] %% activate for two-column option
\noindent

There are several fundamental differences in properties of nonlinear conservative and dissipative systems. One of them is related to the structure of stationary modes. In conservative systems such modes constitute  one- (or several-) parametric families, while in the presence of gain and dissipation stationary solutions represent isolated fixed points (see e.g.~\cite{Akhmed}). This difference stems from the fact that, besides the balance between the nonlinearity and dispersion necessary for existence of any conservative  mode, stationary  modes in a dissipative system must satisfy  an additional constraint of the balance between the dissipation and gain. The latter  constraint,  generally speaking, eliminates one (or even several) free parameters and does not allow for existence of  continuous parametric families which could exist in the conservative case. The described difference has particular importance for nonlinear optics, where the propagation of nonlinear modes is defined by the input signal in the conservative case and by the system parameters in the dissipative case. A possibility of simultaneous manipulating nonlinear modes by both means
%(input signal and system parameters)
would greatly enhance versatility of an optical system.

Recently, this issue attracted special attention in the context
of nonlinear extensions of   parity-time-  ($\PT$-) symmetric~\cite{Bender} systems, whose optical applications were introduced theoretically \cite{Muga,Christ,nonlin_scarf} and implemented experimentally \cite{exper1,exper2}. It  is now established that the $\PT$-symmetric optical potentials, although being complex, i.e. involving  gain and dissipation, still can possess continuous parametric families of the localized modes (resembling in this way   conservative systems). In particular, continuous families of nonlinear modes were obtained for discrete optical systems~\cite{discrete,ZK},  for parabolic~\cite{parab_nonlin1,parab_nonlin2}, periodic~\cite{periodic}, localized~\cite{localized} and double-well \cite{double-well1,double-well2,Jianke2} $\PT$-symmetric potentials with Kerr nonlinearity.  The continuous families of solutions were  also obtained for nonlinear $\PT$-symmetric potentials~\cite{nonlin} and for mixed linear and nonlinear potentials~\cite{mixed}, as well as  for optical  $\PT$-symmetric systems with  $\chi^{(2)}$ nonlinearity, in both continuous~\cite{chi2_cont} and discrete~\cite{chi2_disc} settings.  Models allowing for families of exact solutions were reported for continuous~\cite{famil_cont} and discrete~\cite{famil_disc} media with $\PT$-symmetric defects. It was also conjectured  that the $\PT$ symmetry is  a necessary condition for the existence of  continuous  families of nonlinear modes in one-dimensional potentials  \cite{JY2}.

In this context, an interesting  numerical result was reported in the recent work \cite{FatkOL14}. Within the framework of the parabolic approximation
\begin{eqnarray}
\label{NLS_gen}
i\Psi_z=-\Psi_{xx} - V(x)\Psi-g|\Psi|^2\Psi,
\end{eqnarray}
with  $g=1$ ($g=-1$) corresponding to the focusing (defocusing) nonlinearity, and with the complex potential
\begin{eqnarray}
\label{VW}
V(x) = w^2(x)-iw_x(x),
\end{eqnarray}
the authors
found that for  $w(x)=\eta/\cosh(\alpha x)$, where $\eta$ is a constant,  the system supports continuous families of nonlinear modes even if  $\alpha=\alpha_-$ for $x<0$ and $\alpha=\alpha_+$ for $x\geq 0$ with  $\alpha_- \ne \alpha_+$. The importance of this finding stems from the fact that for $\alpha_- \ne \alpha_+$ the potential (\ref{VW}) is not $\PT$ symmetric, i.e., $V(x) \neq  V^*(-x)$. Notice that if $\alpha$ is  constant in the whole space (i.e., $\alpha_- = \alpha_+$), then potential (\ref{VW}) becomes $\PT$ symmetric and  is usually referred to as Scarf II potential. Its linear spectrum is well known \cite{AkhmedScarff}, and its nonlinear modes at specific values of the parameters were reported in~\cite{nonlin_scarf}.

The asymmetric Scarf II potential considered in \cite{FatkOL14}
was the first example of an asymmetric complex one-dimensional potential  allowing  for continuous families of nonlinear modes. Naturally,
the results of \cite{FatkOL14} raised  several important issues. The first one is the elaboration of analytical arguments which would corroborate the numerical results of \cite{FatkOL14} on the existence of continuous families.  Second, it is highly relevant to understand the nature of such modes and to examine possibilities for   generalization of the reported numerical findings. Third, it is of practical importance to establish whether continuous families persist if instead of an asymmetric one-hump function $w(x)$ (this was an important constraint in \cite{FatkOL14}) one considers more complex (say one- or multi-hump) asymmetric  functions $w(x)$.
Finally, it is important to understand if the  class  of functions $V(x)$ having the form (\ref{VW}) can be extended to a  more general class that preserves the existence of the continuous families.

The main goal of this Letter is to propose answers to the above questions. More specifically, we show that   continuous families in the problem (\ref{NLS_gen})--(\ref{VW}) are explained by a ``hidden'' symmetry, which is expressed in the form of a conserved quantity of the nonlinear dynamical system describing profiles of the nonlinear modes.  This  conserved quantity exists for any real differentiable function $w(x)$, and therefore, {\em the continuous  families exist in complex potentials of a rather general form}. Our finding allows us to develop a  demonstrative computation approach which can be used as a tool for a simple  and reliable numerical calculation of nonlinear modes of this novel type. In order to illustrate our findings and to emphasize  possibilities for generalization of   the previous results,  we introduce an {\em asymmetric complex double-hump potential} and show that it features real linear spectrum and  possesses continuous families of nonlinear modes. The main properties of the found nonlinear modes are also described.

Let us briefly discuss the properties of the  spectrum of potential (\ref{VW}). The idea of using  (\ref{VW}) for constructing $\PT$-symmetric potentials with entirely real spectra was suggested in~\cite{Wadati}. It is based on the existence of  a connection  between the Zakharov-Shabat (ZS) spectral problem
\begin{equation}
\label{ZS}
\phi_{1,x}+i\zeta \phi_1=w(x,\tau)\phi_2, \,\,\, \phi_{2,x}-i\zeta \phi_2=-w(x,\tau)\phi_1,
\end{equation}
associated with the modified Korteweg-de Vries (mKdV) equation, $w_\tau+6w^2w_x+w_{xxx}=0$,  and the   Schr\"{o}dinger eigenvalue problem   $\beta\phi=-\phi_{xx} - V(x,\tau)\phi$ with the potential $V(x)$ given by (\ref{VW}).  The spectral parameters of the two problems are   related as $\beta=-\zeta^2$ (see e.g.~\cite{Wadati,Lamb} for more details).  Presence of this connection  has several important consequences. To formulate them, we recall (see e.g. \cite{Lamb}) that discrete eigenvalues of the ZS problem (\ref{ZS}), if any,  are either purely imaginary or situated symmetrically with respect to the imaginary axis (i.e. if $\zeta$ is an eigenvalue, then $-\zeta^*$ is an eigenvalue, as well), and the continuous spectrum is the real axis. Thus from {\em any} solution $w(x,\tau)$ of the mKdV equation that provides pure imaginary discrete eigenvalues  of (\ref{ZS}), one can obtain a complex potential $V(x,\tau)$, defined by (\ref{VW}) with purely real spectrum $\beta$. Next, we notice that $w(x,\tau)$   depends  on $\tau$  which is time in the mKdV equation [in this paragraph we explicitly indicate this dependence by writing down $w(x,\tau)$ instead of  $w(x)$]. The spectrum of the ZS problem  does not depend on $\tau$. This means that $\tau$ can be considered as a ``deformation'' parameter, and each solution $w(x,\tau)$  represents  a family of deformable potentials $V(x, \tau)$ which can be either $\PT$ symmetric, as   in~\cite{Wadati}, or not, as   in~\cite{FatkOL14},  but in any case having real spectrum. Finally, the constraint $\zeta=-\zeta^*$, imposed on discrete eigenvalues, makes the real spectrum of $V(x,\tau)$ sufficiently robust, i.e. starting with a purely solitonic  solutions, like in~\cite{Wadati}, one can change their shape creating potentials of more sophisticated forms, without violating reality of the spectrum of $V(x)$. In this way, one can construct asymmetric one- and  multi-hump potentials $V(x)$ of the form (\ref{VW}) with purely real spectrum.

Now we turn to stationary nonlinear modes  which are written down in the form $\Psi(x,z)= e^{i \beta z} \psi(x)$, where $\beta$ is the real propagation constant and $\psi(x)$ is the complex field. It is convenient to employ the  ``hydrodynamic'' representation  $\psi(x) = \rho(x)e^{i\theta(x)}$, where $\rho(x)$ is the amplitude and $\theta(x)$ is the phase of the field. Then, introducing the phase gradient $v(x)\equiv\theta_x(x)$, we reduce Eq.~(\ref{NLS_gen}) to the system
\begin{subequations}
    \label{hydro}
\begin{eqnarray}
    \rho_{xx} - \beta \rho + w^2 \rho + g \rho^3 - v^2 \rho = 0,\\
    2\rho_x v + \rho v_x - w_x \rho = 0.
\end{eqnarray}
\end{subequations}
It is convenient to look at these equations as at a nonlinear dynamical system with respect to evolution variable  $x$, i.e. if a solution $(\rho,v)$ is given at some $x_0$, then  its ``evolution'' towards growing or decaying $x$ will describe the transverse profile of the stationary nonlinear mode.

An important finding of our study is that system (\ref{hydro}) has a ``conserved'', i.e. $x$-independent, quantity
\begin{equation}
I = \rho_x^2 + \rho^2(v-w)^2 - \beta \rho^2 +  {g}\rho^4 / 2,\quad  dI/dx\equiv 0.
\end{equation}
The specific form of the potential (\ref{VW}) is crucial for the existence of this conserved quantity: it does not exist in  a system with a generic complex potential $V(x)$.
 The conserved quantity $I$ appears to be crucial for understanding the  existence of continuous families of nonlinear modes. Focusing on spatially localized modes $\psi(x)$ which satisfy the boundary conditions
$\psi(x)\to 0$ as
$|x|\to \infty$,
we conclude that for any localized mode $I=0$, i.e.
\begin{equation}
\label{constr}
\rho_x^2 + \rho^2(v-w)^2 - \beta \rho^2 +  {g} \rho^4/2=0.
\end{equation}
This constraint imposes an additional algebraic relation among the dependent variables $\rho$, $\rho_x$ and $v$, which   allows one to elaborate simple  arguments for  explanation of the  existence of continuous  families of nonlinear modes.

Indeed, for a function $w(x)$ vanishing as $|x|\to\infty$ sufficiently rapidly, the respective asymptotic of the modes reads $\psi  \sim r_+ e^{-\sqrt{\beta} x+i\theta_+}$ as $x\to \infty$ and   $\psi  \sim r_- e^{\sqrt{\beta} x+i\theta_-}$, as $x\to -\infty$,   where $r_\pm$ and $\theta_\pm$ are real constants. Any mode must satisfy the continuity condition for the field and for its derivative at any intermediate point $x_0$~\cite{Landau}, i.e. there are {\em three} matching conditions which must be satisfied at $x_0$: $\rho_-(x_0)= \rho_+(x_0)$, $\rho_{-, x}(x_0)= \rho_{+, x}(x_0)$ and $v_-(x_0)=v_+(x_0)$, where the subscripts ``$-$'' and ``$+$'' stay for the solutions at $x\leq x_0$ and $x\geq x_0$, respectively.  Since the constant phases $\theta_{\pm}$ do  not affect either the amplitude or the phase gradient, for a generic complex potential these matching conditions must be satisfied using only {\em two} constants $r_\pm$, which  is generally not possible unless the propagation constant $\beta$ is adjusted to acquire some particular value. This explains the absence of the continuous families in complex potentials  of a generic type. However, in our case, due to the integral $I$,   there effectively exist only {\em two} matching conditions: if any  two   conditions   are satisfied then the third one is satisfied automatically due to (\ref{constr}). For example, if $\rho_-(x_0)= \rho_+(x_0)$ and $v_-(x_0)=v_+(x_0)$ then Eq.~(\ref{constr}) implies  $[\rho_{-,x}(x_0)]^2 = [\rho_{+,x}(x_0)]^2$. Notice, that in this case Eq.~(\ref{constr}) is satisfied also by $\rho_{-,x}(x_0) = -\rho_{+,x}(x_0)$,  which  is  a ``phantom'' solution   not corresponding to a differentiable nonlinear mode (i.e. it must be discarded). However, in  any practical situation, one can easily check whether the necessary condition  $\rho_{-,x}(x_0) =  \rho_{+,x}(x_0)$ holds, i.e. the artificial ambiguity is easily removed.

Thus we argue  that for an \textit{arbitrary} propagation constant $\beta$ a localized nonlinear mode can be found by solving two matching equations with two real parameters $r_+$ and $r_-$. Hence by varying the free parameter $\beta$ one can construct continuous families of nonlinear modes.

\begin{figure}%[h]
	\centering
	\includegraphics[width=\columnwidth]{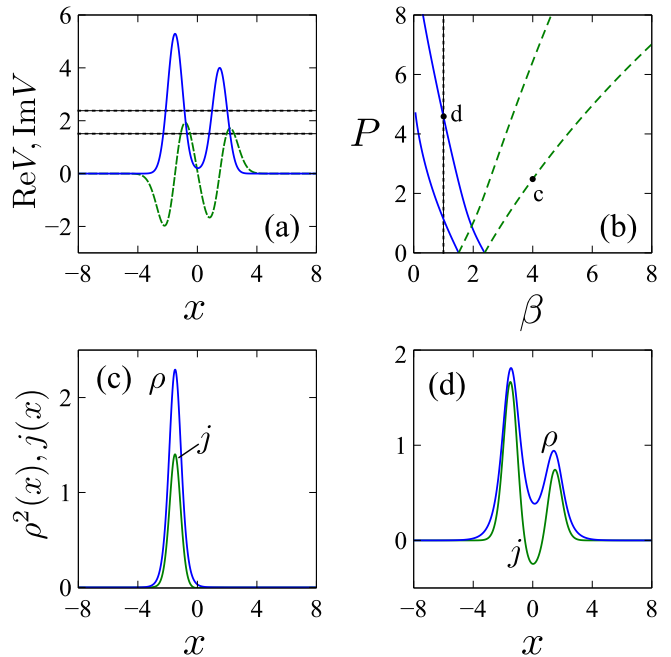}
	\caption{(Color online) (a) Real (blue solid line) and imaginary (green dashed line) parts of the
		%asymmetric double-hump
		potential $V(x)$ for $w(x)$ defined by (\ref{dw}). Horizontal dotted lines show values of the linear eigenvalues $\beta_{1,2}$. (b) Families of localized nonlinear modes. % visualized as dependencies $P$ \textit{vs.} $\beta$.
		Blue solid lines (green dashed lines) correspond to defocusing (focusing) nonlinearity. Points labeled as ``c'' and ``d'' show the nonlinear modes
		% whose spatial profiles are
		presented in panels (c) and (d), respectively. Vertical dotted line at $\beta=1$ indicates the two nonlinear modes found by means of the demonstrative computation in Fig.~\ref{fig2}. (c) The nonlinear mode in focusing medium at $\beta=4$. (d) The nonlinear mode in defocusing medium at $\beta=1$.}
	\label{fig1}
\end{figure}

The described arguments hold for a general class of functions $w(x)$ and allow for simple and illustrative computation of the nonlinear modes.  As an  example, we consider an asymmetric double-hump potential (cf.  the
single-hump potential in \cite{FatkOL14}) choosing
%the following function:
\begin{equation}
\label{dw}
w(x) = a_- e^{-(x+ l)^2} + a_+ e^{-(x-l)^2}.
\end{equation}
The $\PT$-symmetric version  of this potential (with $a_-=a_+$) was recently considered in \cite{Jianke2}.
We  focus  on an {\em asymmetric} case $a_-\neq a_+$ and for numerical illustrations set $a_-=2.3$,  $a_+=2$ and $l=1.5$. Real and imaginary parts of the  resulting asymmetric potential are shown in Fig.~\ref{fig1}(a).  Potential $V(x)$ has two real isolated eigenvalues,  $\beta_1 \approx 1.51$ and $\beta_2 \approx 2.38$, and  the continuous spectrum  lying on the negative  half of the real axis.

\begin{figure}[h]
\centering
\includegraphics[width=0.5\columnwidth]{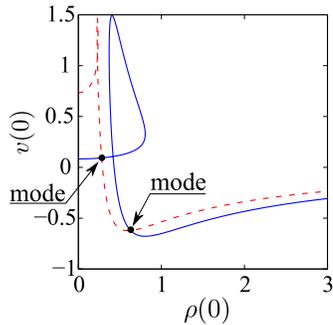}
\caption{(Color online) Demonstrative computation  of nonlinear modes. Solid blue [dashed red] curve shows dependence $v_-(0)$ \textit{vs.}   $\rho_-(0)$  [$v_+(0)$ \textit{vs.}   $\rho_+(0)$] obtained after scanning the range $r_- \in [0, 8.5]$ ($r_+ \in [0, 8.5]$), for $g=-1$ and $\beta=1$. The  intersections between the two curves
	correspond to two nonlinear modes for  $g=-1$  and $\beta=1$, see  also {Fig.~\ref{fig1}(b)}.}
\label{fig2}
\end{figure}

To verify the existence of nonlinear modes, we scanned a certain range  of parameters $r_-$ and $r_+$, and  computed the values $\rho_\pm(0)$,  $ \rho_{\pm, x}(0)$ and $v_{\pm}(0)$, i.e. choosing  $x_0=0$. The obtained dependencies are plotted in Fig.~\ref{fig2} on the plane $(\rho(0),v(0))$.
Any intersection of the solid (blue) and dashed (red) curves corresponds to the moment when  $\rho_-(0)= \rho_+(0)$ and $v_-(0)=v_+(0)$ for certain  $r_+$ and $r_-$. As explained above, this automatically implies  $[\rho_{-,x}(0)]^2 = [\rho_{+,x}(0)]^2$. We  also checked that for both the intersections  $\rho_{-,x}(0) = \rho_{+,x}(0)$, and   therefore they correspond to the two localized modes. The  values of the parameters $r_\pm$  at the crossing points can be  used to compute   spatial profiles of the nonlinear modes.
Notice that in order to establish the existence of the  solutions, significant efforts in the high numerical accuracy were necessary in the previous works (the accuracy of the simulations reported in \cite{FatkOL14} and \cite{Jianke2} was respectively up to $10^{-10}$ and $10^{-30}$). The ``numerical proof'' of the existence of the asymmetric modes presented here does not require any specific concern about the accuracy, as the existence of the modes is established simply by crossing of the two curves, as in Fig.~\ref{fig2}.

Using the described approach for different values of $\beta$, we have identified different families of nonlinear modes in the asymmetric complex double-hump potential, in both focusing ($g=1$) and defocusing ($g=-1$) nonlinearities.  The results  can be  visualized in the form of dependencies of the power flow $P = \int_{-\infty}^\infty |\psi(x)|^2 dx$ on the propagation constant $\beta$ [Fig.~\ref{fig1}(b)]. The families of nonlinear modes presented in Fig.~\ref{fig1}(b)  bifurcate from the isolated eigenvalues $\beta_{1,2}$ of the complex potential $V(x)$. In the case of the  defocusing nonlinearity, the families asymptotically approach the edge of the continuous spectrum, i.e. $\beta=0$.

We checked the linear stability of the  modes, performing the standard substitution $\Psi= e^{i\beta z}[\psi(x) + A(x) e^{\lambda z} + B^*(x) e^{\lambda^* z}]$, $|A|,|B| \ll |\psi|$, and computing instability increment Re$\lambda$  from the linearized problem.
This analysis shows that the families  presented in Fig.~\ref{fig1}(b) are  stable, at least in the explored range of the propagation constant $\beta$. This result  complies with the Vakhitov-Kolokolov (VK) criterion~\cite{VK} for the focusing  and anti-VK criterion~\cite{SM} for the defocusing media. We also notice  that the families shown in Fig.~\ref{fig1}(b) do not exhaust all possible nonlinear modes. For example, in the focusing medium, we also found a family of nonlinear modes bifurcating from the edge of the continuous spectrum, i.e. from $\beta=0$. However, this family is completely unstable and is not presented.

Specific spatial distributions of the field intensity of the modes are illustrated in Fig.~\ref{fig1}(c,d). The  modes feature  highly asymmetric intensity profiles $\rho^2=|\psi(x)|^2$, as well as asymmetric currents $j=v\rho^2$,  which reflects nontrivial power transfer associated with such modes. Generically,  the nonlinearity (either focusing or defocusing) results in deformation of the linear modes without appreciable qualitative changes of their structure.

We also computed the propagation of several nonlinear modes subject to a relatively small ($5\%$ of the amplitude) perturbation. The results (not presented here) indicate that the nonlinear modes are robust against the  introduced perturbation, which corroborates  stability of the found solutions.

To conclude, we have explained and generalized recent numerical findings of Ref.~\cite{FatkOL14}. We have shown that the  continuous families of nonlinear modes in a complex asymmetric potential can be understood by the peculiar property of the underlying dynamical system which possesses a  ``conserved'' (i.e. $x$-independent) quantity. Using this finding, we suggested a simple numerical approach which allows to confirm easily (without resorting to    calculations of extremely high accuracy) the existence of families of nonlinear modes in non-$\PT$-symmetric complex potentials. We have also generalized the previous results on generic potentials allowing the representation  $w^2(x) - i w_x(x)$ and have demonstrated explicitly that the remarkable properties of these complex potentials (i.e.,  reality of the spectrum and existence of the continuous families) persist if instead of a single-hump function $w(x)$ one considers more complex asymmetric double- (or multi-hump)  functions. We expect that the families of localized nonlinear modes should also exist in asymmetric complex periodic potentials.  While our results are obtained for focusing and defocusing cubic nonlinearities,   generalizations for nonlinearities of other types (say quintic or cubic-quintic) are straightforward.

Authors appreciate discussions with  Profs. F. Kh. Abdullaev, J. Yang, and  V. E. Vekslerchik, and acknowledge support of FCT (Portugal) under the grants  PEst-OE/FIS/UI0618/2014 and  PTDC/FIS-OPT/1918/2012.

\pagebreak

\section*{List of references with titles}

\end{document}